\RequirePackage{fix-cm}

\documentclass[twocolumn]{svjour3}          

\smartqed  

\usepackage{graphicx}
\usepackage{natbib}
\usepackage{amsmath}
\usepackage{amssymb}
\usepackage{algorithm}

\newcommand{\dd}{\,d}
\newcommand{\mref}[1]{(\ref{#1})}
\newcommand{\Prob}{\operatorname{P}}

\newcommand{\given}{\operatorname{|}}
\newcommand{\X}{\mathbf{X}}
\newcommand{\G}{\mathcal{G}}
\newcommand{\D}{\mathcal{D}}

\newcommand{\PXi}{\Pi_{X_i}}
\newcommand{\XPi}{X_i \given \PXi}
\newcommand{\T}{\Theta_{X_i}}
\newcommand{\TT}{\T \given \PXi}

\newcommand{\piijk}{\pi_{ik \given j}}
\newcommand{\dXi}{\delta_{X_i}}
\newcommand{\DXi}{\Delta_{X_i}}
\newcommand{\VDXi}{\mathit{Val}(\DXi)}
\newcommand{\GXi}{\Gamma_{X_i}}
\newcommand{\idXi}{d_{X_i}}
\newcommand{\Gmax}{\G_{\mathit{max}}}
\newcommand{\Smax}{S_{\mathit{max}}}
\newcommand{\Score}{\mathit{Score}}

\newcommand{\sumi}{\sum_{i = 1}^{N}}
\newcommand{\sumj}{\sum_{j = 1}^{\idXi + 1}}
\newcommand{\sumk}{\sum_{k = 1}^{N - 1}}

\newcommand{\VAR}{\operatorname{VAR}}
\newcommand{\COV}{\operatorname{COV}}

\def\bbbone{{\mathchoice {\rm 1\mskip-4mu l} {\rm 1\mskip-4mu l}
            {\rm 1\mskip-4.5mu l} {\rm 1\mskip-5mu l}}}

\journalname{Statistics and Computing}

\begin{document}

\title{Learning Bayesian Networks from Big Data with Greedy Search}
\subtitle{Computational Complexity and Efficient Implementation}

\titlerunning{Learning Bayesian Networks from Big Data}

\author{Marco Scutari \and Claudia Vitolo \and Allan Tucker}


\institute{M. Scutari \at
           Department of Statistics, University of Oxford,
           24--29 St. Giles', Oxford OX1 3LB, United Kingdom.
           \email{scutari@stats.ox.ac.uk}
           \and
           C. Vitolo \at
           Forecast Department, European Centre for Medium-range Weather
           Forecast, Reading, United Kingdom.
           \and
           A. Tucker \at
           Department of Computer Science, Brunel University London,
           Kingston Lane, Uxbridge, United Kingdom.
}

\date{Received: \ldots / Accepted: \ldots}

\maketitle

\begin{abstract}
Learning the structure of Bayesian networks from data is known to be a
computationally challenging, NP-hard problem. The literature has long
investigated how to perform structure learning from data containing large
numbers of variables, following a general interest in high-dimensional
applications (``small $n$, large $p$'') in systems biology and genetics.

More recently, data sets with large numbers of observations (the so-called
``big data'') have become increasingly common; and these data sets are not
necessarily high-dimensional, sometimes having only a few tens of variables
depending on the application. We revisit the computational complexity of
Bayesian network structure learning in this setting, showing that the common
choice of measuring it with the number of estimated local distributions leads to
unrealistic time complexity estimates for the most common class of score-based
algorithms, greedy search. We then derive more accurate expressions under common
distributional assumptions. These expressions suggest that the speed of Bayesian
network learning can be improved by taking advantage of the availability of
closed form estimators for local distributions with few parents. Furthermore, we
find that using predictive instead of in-sample goodness-of-fit scores improves
speed; and we confirm that is improves the accuracy of network reconstruction as
well, as previously observed by \citet{engineering}. We demonstrate these
results on large real-world environmental and epidemiological data; and on
reference data sets available from public repositories.

\end{abstract}

\section{Introduction}
\label{sec:intro}

Bayesian networks \citep[BNs; ][]{pearl} are a class of graphical models defined
over a set of random variables $\X = \{X_1, \ldots, X_N\}$, each describing some
quantity of interest, that are associated with the nodes of a directed acyclic
graph (DAG) $\G$. (They are often referred to interchangeably.) Arcs in $\G$
express direct dependence relationships between the variables in $\X$, with
graphical separation in $\G$ implying conditional independence in probability.
As a result, $\G$ induces the factorisation
\begin{equation}
  \label{eq:parents}
  \Prob(\X \given \G, \Theta) = \prod_{i=1}^N \Prob(\XPi, \T),
\end{equation}
in which the joint probability distribution of $\X$ (with parameters $\Theta$)
decomposes in one local distribution for each $X_i$ (with parameters $\T$,
$\bigcup_{\X} \T = \Theta$) conditional on its parents $\PXi$.

While in principle there are many possible choices for the distribution of $\X$,
the literature has focused mostly on three cases. \emph{Discrete BNs}
\citep{heckerman} assume that both $\mathbf{X}$ and the $X_i$ are multinomial
random variables. Local distributions take the form
\begin{align*}
  &\XPi \sim \mathit{Mul}(\piijk),& &\piijk = \Prob(X_i = k \given \PXi = j);
\end{align*}
their parameters are the conditional probabilities of $X_i$ given each
configuration of the values of its parents, usually represented as a conditional
probability table for each $X_i$. \emph{Gaussian BNs} \citep[GBNs;][]{heckerman3}
model $\mathbf{X}$ with a multivariate normal random variable and assume that
the $X_i$ are univariate normals linked by linear dependencies. The parameters
of the local distributions can be equivalently written \citep{weatherburn} as
the partial Pearson correlations $\rho_{X_i, X_j \given \PXi \setminus X_j}$
between $X_i$ and each parent $X_j$ given the other parents; or as the
coefficients $\boldsymbol{\beta}_{X_i}$ of the linear regression model
\begin{align*}
  &X_i = \mu_{X_i} + \PXi\boldsymbol{\beta}_{X_i} + \varepsilon_{X_i},&
  &\varepsilon_{X_i} \sim N(0, \sigma^2_{X_i}),
\end{align*}
so that $\XPi \sim N(\mu_{X_i} + \PXi\boldsymbol{\beta}_{X_i}, \sigma^2_{X_i})$.
Finally, \emph{conditional linear Gaussian BNs} \citep[CLGBNs; \linebreak][]{lauritzen}
combine discrete and continuous random variables in a mixture model:
\begin{itemize}
  \item discrete $X_i$ are only allowed to have discrete parents (denoted
    $\DXi$), are assumed to follow a multinomial distribution parameterised
    with conditional probability tables;
  \item continuous $X_i$ are allowed to have both discrete and continuous parents
    (denoted $\GXi$, $\DXi \cup \GXi = \PXi$), and their local distributions are
    \begin{equation*}
      \XPi \sim N(\mu_{X_i, \dXi} + \Gamma_{X_i}\boldsymbol{\beta}_{X_i, \dXi},
                  \sigma^2_{X_i, \dXi})
    \end{equation*}
    which can be written as a mixture of linear regressions
    \begin{multline*}
      X_i = \mu_{X_i, \dXi} + \Gamma_{X_i}\boldsymbol{\beta}_{X_i, \dXi} + \varepsilon_{X_i, \dXi}, \\
      \varepsilon_{X_i, \dXi} \sim N(0, \sigma^2_{X_i, \dXi}),
    \end{multline*}
    against the continuous parents with one component for each configuration
    $\dXi \in \VDXi$ of the discrete parents. If $X_i$ has no discrete parents,
    the mixture reverts to a single linear regression.
\end{itemize}

Other distributional assumptions, such as mixtures of truncated exponentials
\citep{truncexp} or copulas \citep{copula}, have been proposed in the literature
but have seen less widespread adoption due to the lack of exact conditional
inference and simple closed-form estimators.

The task of learning a BN from a data set $\D$ containing $n$ observations is
performed in two steps:
\begin{equation*}
  \underbrace{\Prob(\G, \Theta \given \D)}_{\text{learning}} =
    \underbrace{\Prob(\G \given \D)}_{\text{structure learning}} \cdot
    \underbrace{\Prob(\Theta \given \G, \D)}_{\text{parameter learning}}.
\end{equation*}
\emph{Structure learning} consists in finding the DAG $\G$ that encodes the
dependence structure of the data, thus maximising $\Prob(\G \given \D)$ or
some alternative goodness-of-fit measure; \emph{parameter learning} consists
in estimating the parameters $\Theta$ given the $\G$ obtained from structure
learning. If we assume parameters in different local distributions are
independent \citep{heckerman}, we can perform parameter learning independently
for each node following \mref{eq:parents} because
\begin{equation*}
  \Prob(\Theta \given \G, \D) = \prod_{i=1}^N \Prob(\TT, \D).
\end{equation*}
Furthermore, if $\G$ is sufficiently sparse each node will have a small
number of parents; and $\XPi$ will have a low-dimensional parameter space,
making parameter learning computationally efficient.

On the other hand, structure learning is well known to be both NP-hard
\citep{nphard} and NP-complete \citep{npcomp}, even under unrealistically
favourable conditions such as the availability of an independence and inference
oracle \citep{nplarge}.\footnote{Interestingly, some relaxations of BN structure
learning are not NP-hard; see for example \citet{notnphard} on learning the
structure of causal networks.} This is despite the fact that if we take
\begin{equation*}
  \Prob(\G \given \D) \propto \Prob(\G)\Prob(\D \given \G),
\end{equation*}
again following \mref{eq:parents} we can decompose the marginal likelihood
$\Prob(\D \given \G)$ into one component for each local distribution
\begin{multline*}
    \Prob(\D \given \G)
    = \int \Prob(\D \given \G, \Theta) \Prob(\Theta \given \G) \dd\Theta = \\
    = \prod_{i=1}^N \int \Prob(\XPi, \T) \Prob(\TT) \dd\T;
\end{multline*}
and despite the fact that each component can be written in closed form for
discrete BNs \citep{heckerman}, GBNs \citep{heckerman3} and \linebreak CLGBNs
\citep{bottcher}. The same is true if we replace $\Prob(\D \given \G)$ with
frequentist goodness-of-fit scores such as BIC \citep{schwarz}, which is
commonly used in structure learning because of its simple expression:
\begin{equation*}
  \mathrm{BIC}(\G, \Theta \given \D) = \sumi \log \Prob(\XPi, \T) - \frac{\log(n)}{2} |\T|.
\end{equation*}
Compared to marginal likelihoods, BIC has the advantage that it does not depend
on any hyperparameter, while converging to $\log\Prob(\D \given \G)$
as $n \to \infty$.

These score functions, which we will denote with $\Score(\G, \D)$ in the
following, have two important properties:
\begin{itemize}
  \item they decompose into one component for each local distribution following
    \mref{eq:parents}, say
    \begin{equation*}
      \Score(\G, \D) = \sumi \Score(X_i, \PXi, \D),
    \end{equation*}
    thus allowing local computations (\emph{decomposability});
  \item they assign the same score value to DAGs that encode the same probability
    distributions, and can therefore be grouped in an \emph{equivalence classes}
    \citep[\emph{score equivalence};][]{chickering}.\footnote{All DAGs in the
    same equivalence class have the same underlying undirected graph and
    v-structures (patterns of arcs like $X_i \rightarrow X_j \leftarrow X_k$,
    with no arcs between $X_i$ and $X_k$).}
\end{itemize}
Structure learning via score maximisation is performed using general-purpose
optimisation techniques, typically heuristics, adapted to take advantage of
these properties to increase the speed of structure learning. The most common
are \emph{greedy search} strategies that employ local moves designed to affect
only few local distributions, to that new candidate DAGs can be scored without
recomputing the full $\Prob(\D \given \G)$. This can be done either in the space
of the DAGs with hill-climbing and tabu search \citep{norvig}, or in the space
of the equivalence classes with Greedy Equivalent Search \citep[GES;][]{ges}.
Other options that have been explored in the literature are genetic algorithms
\citep{genetic} and ant colony optimisation \citep{ant}. Exact maximisation of
$\Prob(\D \given \G)$ and BIC has also become feasible for small data sets in
recent years thanks to increasingly efficient pruning of the space of the DAGs
and tight bounds on the scores \citep{cutting,suzuki17,scanagatta}.

In addition, we note that it is also possible to perform structure learning
using conditional independence tests to learn conditional independence
constraints from $\D$, and thus identify which arcs should be included in $\G$.
The resulting algorithms are called \emph{constraint-based algorithms}, as
opposed to the \emph{score-based algorithms} we introduced above; for an
overview and a comparison of these two approaches see \citet{crc13}.
\citet{nplarge} proved that constraint-based algorithms are also NP-hard for
unrestricted DAGs; and they are in fact equivalent to score-based algorithms
given a fixed topological ordering when independence constraints are tested with
statistical tests related to cross-entropy \citep{cowell}. For these reasons, in
this paper we will focus only on score-based algorithms while recognising that a
similar investigation of constraint-based algorithms represents a promising
direction for future research.

The contributions of this paper are:
\begin{enumerate}
  \item to provide general expressions for the (time) computational complexity
    of the most common class of score-based structure learning algorithms,
    \emph{greedy search}, as a function of the number of variables $N$, of the
    sample size $n$, and of the number of parameters $|\Theta|$;
  \item to use these expressions to identify two simple yet effective
    optimisations to speed up structure learning in ``big data'' settings in
    which $n \gg N$.
\end{enumerate}
Both are increasingly important when using BNs in modern machine learning
applications, as data sets with large numbers of observations (the so-called
``big data'') are becoming as common as classic high-dimensional data
(``small $n$, large $p$'', or ``small $n$, large $N$'' using the notation
introduced above). The vast majority of complexity and scalability results
\citep{pcalgo,scanagatta} and computational optimisations \citep{jss14} in
the literature are derived in the latter setting and implicitly assume
$n \ll N$; they are not relevant in the former setting in which $n \gg N$.
Our contributions also complement related work on advanced data structures for
machine learning applications, which include ADtrees \citep{moore}, frequent
sets \citep{goldenberg} and more recently bitmap representations combined with
radix sort \citep{jaroslaw}. Such literature focuses on discrete variables,
whereas we work in a more general setting in which data can include both
discrete and continuous variables.

The material is organised as follows. In Section \ref{sec:complexity} we will
present in detail how greedy search can be efficiently implemented thanks to the
factorisation in \mref{eq:parents}, and we will derive its computational
complexity as a function $N$; this result has been mentioned in many places in
the literature, but to the best of our knowledge its derivation has not been
described in depth. In Section \ref{sec:revisiting} we will then argue that the
resulting expression does not reflect the actual computational complexity of
structure learning, particularly in a ``big data'' setting where $n \gg N$; and
we will re-derive it in terms of $n$ and $|\Theta|$ for the three classes of BNs
described above. In Section \ref{sec:bigdata} we will use this new expression to
identify two optimisations that can markedly reduce the overall computational
complexity of learning GBNs and CLGBNs by leveraging the availability of closed
form estimates for the parameters of the local distributions and out-of-sample
goodness-of-fit scores. Finally, in Section \ref{sec:sims} we will demonstrate
the improvements in speed produced by the proposed optimisations on simulated
and real-world data, as well as their effects on the accuracy of learned
structures.

\section{Computational Complexity of Greedy Search}
\label{sec:complexity}

\begin{algorithm}[t]
\caption{Greedy Search}
\label{algo:greedy}
\vspace{2mm}
\textbf{Input:} a data set $\D$ from $\X$, an initial DAG $\G$ (usually
  the empty DAG), a score function $\Score(\G, \D)$.\\
\textbf{Output:} the DAG $\Gmax$ that maximises $\Score(\G, \D)$.
\begin{enumerate}
  \item Compute the score of $\G$, $S_{\G} = \Score(\G, \D)$.
    \label{step:setup1}
  \item Set $\Smax = S_{\G}$ and $\Gmax = \G$.
    \label{step:setup2}
  \item \textbf{Hill climbing:} repeat as long as $\Smax$ increases:
  \label{step:hc}
  \begin{enumerate}
    \item for every possible arc addition, deletion or reversal in $\Gmax$
      resulting in a DAG:
      \begin{enumerate}
        \item compute the score of the modified DAG $\G^*$,
          $S_{\G^*} = \Score(\G^*, \D)$:
        \item if $S_{\G^*} > \Smax$ and $S_{\G^*} > S_{\G}$, set $\G = \G^*$
          and $S_{\G} = S_{\G^*}$.
      \end{enumerate}
      \item if $S_\G > S_{\mathit{max}}$, set $\Smax = S_\G$ and $\Gmax = \G$.
  \end{enumerate}
  \item \textbf{Tabu search:} for up to $t_0$ times: \label{step:tabu}
  \begin{enumerate}
    \item repeat step \ref{step:hc} but choose the DAG $\G$ with the highest
      $S_\G$ that has not been visited in the last $t_1$ steps regardless of
      $\Smax$;
    \item if $S_\G > \Smax$, set $S_0 = \Smax = S_\G$ and $\G_0 = \Gmax = \G$
      and restart the search from step \ref{step:hc}.
  \end{enumerate}
  \item \textbf{Random restart:} for up to $r$ times, perturb $\Gmax$ with
    multiple arc additions, deletions and reversals to obtain a new DAG $\G'$
    and: \label{step:restart}
    \begin{enumerate}
      \item set $S_0 = \Smax = S_\G$ and $\G_0 = \Gmax = \G$ and restart the
        search from step \ref{step:hc};
      \item if the new $\Gmax$ is the same as the previous $\Gmax$, stop and
        return $\Gmax$.
    \end{enumerate}
\end{enumerate}
\end{algorithm}

A state-of-the-art implementation of greedy search in the context of BN
structure learning is shown in Algorithm \ref{algo:greedy}. It consists of an
initialisation phase (steps \ref{step:setup1} and \ref{step:setup2}) followed
by a \emph{hill climbing} search (step \ref{step:hc}), which is then
optionally refined with \emph{tabu search} (step \ref{step:tabu}) and
\emph{random restarts} (step \ref{step:restart}). Minor variations of this
algorithm have been used in large parts of the literature on BN structure
learning with score-based methods \citep[some notable examples
are][]{heckerman,mmhc,sem}.

Hill climbing uses local moves (arc additions, deletions and reversals) to
explore the neighbourhood of the current candidate DAG $\Gmax$ in the space of
all possible DAGs in order to find the DAG $\G$ (if any) that increases the
score $\Score(\G, \D)$ the most over $\Gmax$. That is, in each iteration hill
climbing tries to delete and reverse each arc in the current optimal DAG $\Gmax$;
and to add each possible arc that is not already present in $\Gmax$. For all
the resulting DAGs $\G^*$ that are acyclic, hill climbing then computes
$S_{\G^*} = \Score(\G^*, \D)$; cyclic graphs are discarded. The $\G^*$ with the
highest $S_{\G^*}$ becomes the new candidate DAG $\G$. If that DAG has a score
$S_{\G} > \Smax$ then $\G$ becomes the new $\Gmax$, $\Smax$ will be set to
$S_{\G}$, and hill climbing will move to the next iteration.

This greedy search eventually leads to a DAG $\Gmax$ that has no neighbour with
a higher score. Since hill climbing is an optimisation heuristic, there is no
theoretical guarantee that $\Gmax$ is a global maximum. In fact, the space of
the DAGs grows super-exponentially in $N$ \citep{harary}; hence multiple local
maxima are likely present even if the sample size $n$ is large. The problem
may be compounded by the existence of score-equivalent DAGs, which by definition
have the same $S_{\G}$ for all the $\G$ falling in the same equivalence class.
However, \citet{gillispie} have shown that while the number of equivalence
classes is of the same order of magnitude as the space of the DAGs, most
contain few DAGs and as many as $27.4\%$ contain just a single DAG. This
suggests that the impact of score equivalence on hill climbing may be limited.
Furthermore, greedy search can be easily modified into GES to work directly in
the space of equivalence classes by using different set of local moves,
side-stepping this possible issue entirely.

In order to escape from local maxima, greedy search first tries to move away
from $\Gmax$ by allowing up to $t_0$ additional local moves. These moves
necessarily produce DAGs $\G^*$ with $S_{\G*} \leqslant \Smax$; hence the new
candidate DAGs are chosen to have the highest $S_\G$ even if $S_\G < \Smax$.
Furthermore, DAGs that have been accepted as candidates in the last
$t_1$ iterations are kept in a list (the \emph{tabu list}) and are not
considered again in order to guide the search towards unexplored regions of
the space of the DAGs. This approach is called \emph{tabu search} (step
\ref{step:tabu}) and was originally proposed by \citet{glover}. If a new DAG
with a score larger than $\Gmax$ is found in the process, that DAG is taken
as the new $\Gmax$ and greedy search returns to step \ref{step:hc}, reverting
to hill climbing.

If, on the other hand, no such DAG is found then greedy search tries again to
escape the local maximum $\Gmax$ for $r_0$ times with random non-local moves,
that is, by moving to a distant location in the space of the DAGs and starting
the greedy search again; hence the name \emph{random restart} (step
\ref{step:restart}). The non-local moves are typically determined by applying
a batch of $r_1$ randomly-chosen local moves that substantially alter $\Gmax$.
If the DAG that was perturbed was indeed the global maximum, the assumption is
that this second search will also identify it as the optimal DAG, in which
case the algorithm terminates.

We will first study the (time) computational complexity of greedy search under
the assumptions that are commonly used in the literature
\citep[see, for instance, ][]{mmhc,spirtes} for this purpose:
\begin{enumerate}
  \item We treat the estimation of each local distribution as an atomic $O(1)$
    operation; that is, the (time) complexity of structure learning is measured
    by the number of estimated local distributions.
  \item Model comparisons are assumed to always pick the right model, which
    happens asymptotically for \linebreak $n \to \infty$ since marginal
    likelihoods and BIC are globally and locally consistent \citep{ges}.
  \item The true DAG $\G_{REF}$ is sparse and contains $O(cN)$ arcs, where
    $c$ is typically assumed to be between $1$ and $5$.
\end{enumerate}
In steps \ref{step:setup1} and \ref{step:setup2}, greedy search computes all
the $N$ local distributions for $\G_0$. In step \ref{step:hc}, each iteration
tries all possible arc additions, deletions and reversals. Since there are
${N \choose 2}$ possible arcs in a DAG with $N$ nodes, this requires $O(N^2)$
model comparisons. If we assume $\G_0$ is the empty DAG (that is, a DAG with no
arcs), hill climbing will gradually add all the arcs in $\G_{REF}$, one in each
iteration. Assuming $\G_{REF}$ is sparse, and assuming that arcs are removed or
reversed a negligible number of times, the overall computational complexity of
hill climbing is then $O(cN^3)$ model comparisons. Step \ref{step:tabu} performs
$t_0$ more iterations, and is therefore $O(t_0 N^2)$. Therefore, the combined
time complexity of steps \ref{step:hc} and \ref{step:tabu} is $O(cN^3 + t_0 N^2)$.
Each of the random restarts involves changing $r_1$ arcs, and thus we can expect
that it will take $r_1$ iterations of hill climbing to go back to the same
maximum, followed by tabu search; and that happens for $r_0$ times. Overall,
this adds $O(r_0 (r_1N^2 + t_0 N^2))$ to the time complexity, resulting in an
overall complexity $g(N)$ of
\begin{align}
  O(g(N)) &= O(cN^3 + t_0 N^2 + r_0 (r_1N^2 + t_0 N^2)) \notag \\
          &= O(cN^3 + (t_0 + r_0(r_1 + t_0)) N^2).
  \label{eq:naive-bigo}
\end{align}
The leading term is $O(cN^3)$ for some small constant $c$, making greedy
search cubic in complexity.

Fortunately, the factorisation in \mref{eq:parents} makes it possible to
recompute only one or two local distributions for each model comparison:
\begin{itemize}
  \item Adding or removing an arc only alters one parent set; for instance,
    adding $X_j \to X_i$ means that $\PXi = \PXi \cup X_j$, and therefore
    $\Prob(\XPi)$ should be updated to $\Prob(\XPi \cup X_j)$. All the other
    local distributions $\Prob(X_k \given \Pi_{X_j}), X_k \neq X_i$ are
    unchanged.
  \item Reversing an arc $X_j \to X_i$ to $X_i \to X_j$ means that
    $\PXi = \PXi \setminus X_j$ and $\Pi_{X_j} = \Pi_{X_j} \cup X_i$, and so
    both $\Prob(\XPi)$ and $\Prob(X_j \given \Pi_{X_j})$ should be updated.
\end{itemize}
Hence it is possible to dramatically reduce the computational complexity of
greedy search by keeping a cache of the score values of the $N$ local
distributions for the current $\Gmax$
\begin{equation*}
  B_i = \Score_{\mathit{max}}(X_i, \PXi^{\mathit{max}}, \D);
\end{equation*}
and of the $N^2 - N$ score differences
\begin{multline*}
  \Delta_{ij} = \Smax - S_{\G^*} = \\ =
    \Score_{\mathit{max}}(X_i, \PXi^{\mathit{max}}, \D) -
    \Score_{\G^*}(X_i, \PXi^{\G^*}, \D), i \neq j,
\end{multline*}
where $\PXi^{\mathit{max}}$ and $\PXi^{\G^*}$ are the parents
of $X_i$ in $\Gmax$ and in the $\G^*$ obtained by removing (if present) or adding
(if not) $X_j \to X_i$ to $\Gmax$. Only $N$ (for arc additions and deletions) or
$2N$ (for arc reversals) elements of $\Delta$ need to be actually computed in
each iteration; those corresponding to the variable(s) whose parent sets were
changed by the local move that produced the current $\Gmax$ in the previous
iteration. After that, all possible arc additions, deletions and reversals can
be evaluated without any further computational cost by adding or subtracting the
appropriate $\Delta_{ij}$ from the $B_i$. Arc reversals can be handled as a
combination of arc removals and additions (\emph{e.g.} reversing $X_i \to X_j$
is equivalent to removing $X_i \to X_j$ and adding  $X_j \to X_i$). As a result,
the overall computational complexity of greedy search reduces from $O(cN^3)$ to
$O(cN^2)$. Finally, we briefly note that score equivalence may allow further
computational saving because many local moves will produce new $\G^*$ that are
in the same equivalence class as $\Gmax$; and for those moves necessarily
$\Delta_{ij} = 0$ (for arc reversals) or $\Delta_{ij} = \Delta_{ji}$ (for adding
or removing $X_i \to X_j$ and $X_j \to X_i$).

\section{Revisiting Computational Complexity}
\label{sec:revisiting}

In practice, the computational complexity of estimating a local distribution
$\Prob(\XPi)$ from data depends on three of factors:
\begin{itemize}
  \item the characteristics of the data themselves (the sample size $n$, the
    number of possible values for categorical variables);
  \item the number of parents of $X_i$ in the DAG, that is, $|\PXi|$;
  \item the distributional assumptions on $\Prob(\XPi)$, which determine the
    number of parameters $|\T|$.
\end{itemize}

\subsection{Computational Complexity for Local Distributions}
\label{sec:revisiting-local}

If $n$ is large, or if $|\T|$ is markedly different for different $X_i$,
different local distributions will take different times to learn, violating the
$O(1)$ assumption from the previous section. In other words, if we denote the
computational complexity of learning the local distribution of $X_i$ as
$O(f_{\PXi}(X_i))$, we find below that $O(f_{\PXi}(X_i)) \neq O(1)$.

\subsubsection{Nodes in Discrete BNs}
\label{sec:revisiting-dbn-nodes}

In the case of discrete BNs, the conditional probabilities $\piijk$ associated
with each $\XPi$ are computed from the corresponding counts $n_{ijk}$ tallied
from $\{X_i, \PXi\}$; hence estimating them takes $O(n(1 + |\PXi|))$ time.
Computing the marginals counts for each configuration of $\PXi$ then takes
$O(|\T|)$ time; assuming that each discrete variable takes at most $l$ values,
then $|\T| \leqslant l^{1 + |\PXi|}$ leading to
\begin{equation}
  O(f_{\PXi}(X_i)) = O\left(n(1 + |\PXi|) + l^{1 + |\PXi|}\right).
  \label{eq:dbn-nbigo}
\end{equation}

\subsubsection{Nodes in GBNs}
\label{sec:revisiting-gbn-nodes}

In the case of GBNs, the regressions coefficients for $\XPi$ are usually
computed by applying a QR decomposition to the augmented data matrix
$[1 \, \PXi]$:
\begin{align*}
  &[1 \, \PXi] = \mathbf{QR}& &\text{leading to}&
  &\mathbf{R}[\mu_{X_i}, \boldsymbol{\beta}_{X_i}] = \mathbf{Q}^T X_i
\end{align*}
which can be solved efficiently by backward substitution since $\mathbf{R}$ is
upper-triangular. This approach is the \emph{de facto} standard approach for
fitting linear regression models because it is numerically stable even in the
presence of correlated $\PXi$ \citep[see][for details]{seber}. Afterwards we
can compute the fitted values
$\hat{x}_i = \PXi \hat{\boldsymbol{\beta}}_{X_i}$ and the residuals
$X_i - \hat{x}_i$ to estimate
$\hat{\sigma}^2_{X_i} \propto (X_i - \hat{x}_i)^T (X_i - \hat{x}_i)$.
The overall computational complexity is
\begin{align}
  O(&f_{\PXi}(X_i)) = \notag \\
    &= \underbrace{O\left(n(1 + |\PXi|)^2\right)}_{\text{QR decomposition}} +
    \underbrace{O\left(n(1 + |\PXi|\right))}_{\text{computing $\mathbf{Q}^T X_i$}} + \notag \\
    &\qquad\underbrace{O\left((1 + |\PXi|)^2\right)}_{\text{backwards substitution}} +
    \underbrace{O\left(n(1 + |\PXi|)\right)}_{\text{computing $\hat{x}_i$}} + \notag \\
    &\qquad\underbrace{O\left(3n\right)}_{\text{computing $\hat{\sigma}^2_{X_i}$}}
\label{eq:gbn-node}
\end{align}
with leading term $O((n + 1)(1 + |\PXi|)^2)$.

\subsubsection{Nodes in CLGBNs}
\label{sec:revisiting-clgbn-nodes}

As for CLGBNs, the local distributions of discrete nodes are estimated in the
same way as they would be in a discrete BN. For Gaussian nodes, a regression of
$X_i$ against the continuous parents $\GXi$ is fitted from the $n_{\dXi}$
observations corresponding to each configuration of the discrete parents $\DXi$.
Hence the overall computational complexity is
\begin{align}
  O(&f_{\PXi}(X_i)) = \notag \\
    &= \sum_{\dXi \in \VDXi} O\left(n_{\dXi}(1 + |\GXi|)^2\right) + \notag \\
    &\qquad\qquad  O\left(2n_{\dXi}(1 + |\GXi|)\right) + O\left(1 + |\GXi|)^2\right) + \notag \\
    &\qquad\qquad  O\left(3n_{\dXi}\right)\notag \\
    &= O\left(n(1 + |\GXi|)^2\right) + O\left(2n(1 + |\GXi|)\right) + \notag \\
    &\qquad\qquad  O\left(|\VDXi|(1 + |\GXi|)^2\right) + O\left(3n\right) \notag \\
    &= O\left((n + l^{|\DXi|})(1 + |\GXi|)^2\right) + \notag \\
    &\qquad\qquad O\left(2n(1 + |\GXi|)\right) + O\left(3n\right)
\label{eq:clgbn-node}
\end{align}
with leading term $O\left((n + l^{|\DXi|})(1 + |\GXi|)^2\right)$. If $X_i$ has
no discrete parents then \mref{eq:clgbn-node} simplifies to \mref{eq:gbn-node}
since $|\VDXi| = 1$ and $n_{\dXi} = n$.

\subsection{Computational Complexity for the Whole BN}

Let's now assume without loss of generality that the dependence structure of
$\X$ can be represented by a DAG $\G$ with in-degree sequence $d_{X_1} \leqslant
d_{X_2} \leqslant \ldots \leqslant d_{X_N}$. For a sparse graph containing $cN$
arcs, this means $\sumi \idXi = cN$. Then if we make the common choice of
starting greedy search from the empty DAG, we can rewrite \mref{eq:naive-bigo} as
\begin{align}
  O(g(N)) &= O(cN^2) \notag \\
    &= O\left(\sumi \sumj \sumk 1\right) \notag \\
    &= \sumi \sumj \sumk O(1) = O(g(N, \mathbf{d}))
\label{eq:general}
\end{align}
because:
\begin{itemize}
  \item parents are added sequentially to each of the $N$ nodes;
  \item if a node $X_i$ has $\idXi$ parents then greedy search will perform
    $\idXi + 1$ passes over the candidate parents;
  \item for each pass, $N - 1$ local distributions will need to be relearned as
    described in Section \ref{sec:complexity}.
\end{itemize}
The candidate parents in the ($\idXi + 1$)th pass are evaluated but not included
in $\G$, since no further parents are accepted for a node after its parent set
$\PXi$ is complete. If we drop the assumption from Section \ref{sec:complexity}
that each term in the expression above is $O(1)$, and we substitute it with the
computational complexity expressions we derived above in this section, then we
can write
\begin{equation*}
  O(g(N, \mathbf{d}))
    = \sumi \sumj \sumk O(f_{jk}(X_i)).
\end{equation*}
where $O(f_{jk}(X_i)) = O(f_{\PXi^{(j - 1)} \cup X_k}(X_i))$, the computational
complexity of learning the local distribution of $X_i$ conditional of $j - 1$
parents $\PXi^{(j)}$ currently in $\G$ and a new candidate parent $X_k$.

\subsubsection{Discrete BNs}
\label{sec:dbn}

For discrete BNs, $f_{jk}(X_i)$ takes the form shown in \mref{eq:dbn-nbigo} and
\begin{align*}
  O(&g(N, \mathbf{d})) = \\
    &= \sumi \sumj \sumk O(n(1 + j) + l^{1 + j}) \\
    &= O\left(n(c + 1)(N - 1)N + n(N - 1)\sumi \sumj j + \right. \\
    &\hspace{0.52\columnwidth}\left. (N - 1)\sumi \sumj l^{1 + j}\right) \\
    &\approx O\left(ncN^2 + nN \sumi \sumj j +
      N\sumi \sumj l^{1 + j}\right)
\end{align*}
The second term is an arithmetic progression,
\begin{equation*}
  \sumj j = \frac{(\idXi + 1)(\idXi + 2)}{2};
\end{equation*}
and the third term is a geometric progression
\begin{equation*}
  \sumj l^{1 + j} = l^2 \sumj l^{j - 1}
    = l^2 \frac{l^{\idXi + 1} - 1}{l - 1}
\end{equation*}
leading to
\begin{multline}
  O(g(N, \mathbf{d})) \approx \\
     O\left(ncN^2 + nN\sumi \frac{\idXi^2}{2} +
     Nl^2 \sumi \frac{l^{\idXi + 1} - 1}{l - 1} \right).
\label{eq:dbn-whole}
\end{multline}
Hence, we can see that $O(g(N, \mathbf{d}))$ increases linearly in the sample
size. If $\G$ is uniformly sparse, all $\idXi$ are bounded by a constant $b$
($\idXi \leqslant b$, $c \leqslant b$) and
\begin{multline*}
  O(g(N, \mathbf{d})) \approx
    O\left(N^2\left[nc + n\frac{b^2}{2} + l^2\frac{l^{b + 1} - 1}{l - 1}\right]\right),
\end{multline*}
so the computational complexity is quadratic in $N$. Note that this is a stronger
sparsity assumption than \linebreak $\sumi \idXi = cN$, because it bounds
individual $\idXi$ instead of their sum; and it is commonly used to make
challenging learning problems feasible \citep[\emph{e.g.}][]{cooper,friedman}.
If, on the other hand, $G$ is dense and $\idXi = O(N)$, then $c = O(N)$
\begin{equation*}
  O(g(N, \mathbf{d})) \approx O\left(N^2\left[nc + n\frac{N^3}{2} +
    l^2\frac{l^N - 1}{l - 1}\right]\right)
\end{equation*}
and $O(g(N, \mathbf{d}))$ is more than exponential in $N$. In between these two
extremes, the distribution of the $\idXi$ determines the actual computational
complexity of \linebreak greedy search for a specific types of structures. For
instance, if $\G$ is a scale-free DAG \citep{scale-free} the in-degree of most
nodes will be small and we can expect a computational complexity closer to
quadratic than exponential if the probability of large in-degrees decays quickly
enough compared to $N$.

\subsubsection{GBNs}
\label{sec:gbn}

If we consider the leading term of \mref{eq:gbn-node}, we obtain the following
expression:
\begin{align*}
  &O(g(N, \mathbf{d})) = \\
    &= \sumi \sumj \sumk O((n + 1)(j+1)^2) \\
    &= O\left((n + 1)(N - 1) \sumi \sumj (j+1)^2\right) \\
\end{align*}
Noting the arithmetic progression
\begin{equation*}
  \sumj (j+1)^2 = \frac{2\idXi^3 + 15\idXi^2 + 37\idXi + 24}{6}
\end{equation*}
we can the write
\begin{equation*}
  O(g(N, \mathbf{d})) \approx O\left(nN \sumi \frac{\idXi^3}{3} \right),
\end{equation*}
which is again linear in $n$ but cubic in the $\idXi$. We note, however, that
even for dense networks ($\idXi = O(N)$) computational complexity remains
polynomial
\begin{align*}
  O(g(N, \mathbf{d})) \approx O\left(nN^2 \frac{N^3}{3} \right)
\end{align*}
which was not the case for discrete BNs. If, on the other hand $\idXi \leqslant b$,
\begin{align*}
  O(g(N, \mathbf{d})) \approx O\left(nN^2 \frac{b^3}{3} \right)
\end{align*}
which is quadratic in $N$.

\subsubsection{CLGBNs}
\label{sec:clgbn}

Deriving the computational complexity for CLGBNs is more complicated because of
the heterogeneous nature of the nodes. If we consider the leading term of
\mref{eq:clgbn-node} for a BN with $M < N$ Gaussian nodes and $N - M$
multinomial nodes we have
\begin{multline*}
  O(g(N, \mathbf{d})) =
    \sum_{i = 1}^{N - M} \sumj \sum_{k = 1}^{N - M - 1} O(f_{jk}(X_i)) + \\
    \sum_{i = 1}^{M} \sumj \sum_{k = 1}^{N - 1} O(f_{jk}(X_i)).
\end{multline*}
The first term can be computed using \mref{eq:dbn-whole} since discrete nodes
can only have discrete parents, and thus cluster in a subgraph of $N - M$ nodes
whose in-degrees are completely determined by other discrete nodes; and the same
considerations we made in Section \ref{sec:dbn} apply.

As for the second term, we will first assume that all $D_i$ discrete parents of
each node are added first, before any of the $G_i$ continuous parents
($\idXi = D_i + G_i$). Hence we write
\begin{align*}
  &\sum_{i = 1}^{M} \sumj \sum_{k = 1}^{N - 1} O(f_{jk}(X_i)) = \\
    &= \sum_{i = 1}^{M} \left[
         \sum_{j = 1}^{D_i} \sum_{k = 1}^{N - 1} O(f_{jk}(X_i)) +
         \sum_{j = D_i + 1}^{\idXi + 1} \sum_{k = 1}^{N - 1} O(f_{jk}(X_i))
       \right].
\end{align*}
We further separate discrete and continuous nodes in the summations over the
possible $N - 1$ candidates for inclusion or removal from the current parent
set, so that substituting \mref{eq:clgbn-node} we obtain
\begin{align*}
  &\sum_{j = 1}^{D_i} \sum_{k = 1}^{N - 1} O(f_{jk}(X_i)) = \\
    &= \sum_{j = 1}^{D_i} \left[ \sum_{k = 1}^{N - M} O(f_{jk}(X_i)) +
       \sum_{k = 1}^{M - 1} O(f_{jk}(X_i)) \right] \\
    &= \sum_{j = 1}^{D_i} \left[ (N - M) O\left(n + l^j\right) +
       (M - 1) O\left(4\left(n + l^j\right)\right) \right] \\
    &\approx O\left((N + 3M) \sum_{j = 1}^{D_i} \left(n + l^j\right) \right) \\
    &= O\left((N + 3M) \left(nD_i + l \frac{l^{D_i} - 1}{l - 1}\right) \right)
\end{align*}
\begin{align*}
  &\sum_{j = D_i + 1}^{\idXi + 1} \sum_{k = 1}^{N - 1} O(f_{jk}(X_i)) = \\
    &= \sum_{j = D_i + 1}^{\idXi + 1} \left[ \sum_{k = 1}^{N - M} O(f_{jk}(X_i)) +
       \sum_{k = 1}^{M - 1} O(f_{jk}(X_i)) \right] \\
    &= \sum_{j = 1}^{G_i} \Big[ (N - M) O\left(n + l^{D_i}\right) + \\
    &\qquad\qquad (M - 1) O\left(\left(n + l^{D_i}\right)(1 + j)^2\right) \Big] \\
    &\approx O\left(\left(n + l^{D_i}\right) \left( G_i(N - M) + M \frac{G_i^3}{3}\right) \right). \\
\end{align*}
Finally, combining all terms we obtain the following expression:
\begin{align*}
  O(&g(N, \mathbf{d})) \approx \\
    &\approx
     O\left(nc(N - M)^2 + n(N - M)\sum_{i = 1}^{N - M} \frac{\idXi^2}{2} + \right.\\
    &\qquad\qquad \left. (N - M)l^2 \sum_{i = 1}^{N - M} \frac{l^{\idXi + 1} - 1}{l - 1} \right) + \\
    &\qquad \sum_{i = 1}^{M} O\left((N + 3M) \left(nD_i + l \frac{l^{D_i} - 1}{l - 1}\right) \right) + \\
    &\qquad \sum_{i = 1}^{M} O\left(\left(n + l^{D_i}\right) \left( G_i(N - M) + M \frac{G_i^3}{3}\right) \right).
\end{align*}

While it is not possible to concisely describe the behaviour resulting from
this expression given the number of data-dependent parameters ($D_i$, $G_i$,
$M$), we can observe that:
\begin{itemize}
  \item $O(g(N, \mathbf{d}))$ is always linear in the sample size;
  \item unless the number of discrete parents is bounded for both discrete and
    continuous nodes, $O(g(N, \mathbf{d}))$ is again more than exponential;
  \item if the proportion of discrete nodes is small, we can assume that
    $M \approx N$ and $O(g(N, \mathbf{d}))$ is always polynomial.
\end{itemize}

\section{Greedy Search and Big Data}
\label{sec:bigdata}

In Section \ref{sec:revisiting} we have shown that the computational complexity
of greedy search scales linearly in $n$, so greedy search is efficient in the
sample size and it is suitable for learning BNs from big data. However, we have
also shown that different distributional assumptions on $\X$ and on the $\idXi$
lead to different complexity estimates for various types of BNs. We will now
build on these results to suggest two possible improvements to speed up greedy
search.

\subsection{Speeding Up Low-Order Regressions in GBNs and CLGBNs}
\label{sec:closed}

Firstly, we suggest that estimating local distributions with few parents can
be made more efficient; if we assume that $\G$ is sparse, those make up the
majority of the local distributions learned by greedy search and their
estimation can potentially dominate the overall computational cost of Algorithm
\ref{algo:greedy}. As we can see from the summations in \mref{eq:general}, the
overall number of learned local distributions with $j$ parents is
\begin{equation}
  \sumi \bbbone_{\{\idXi \geqslant j - 1\}}(j) = N - \sumi \bbbone_{\{\idXi < j - 1\}}(j),
\label{eq:decay}
\end{equation}
that is, it is inversely proportional to the number of nodes for which $\idXi$
is less than $j- 1$ in the DAG we are learning. If that subset of nodes
represents large fraction of the total, as is the case for scale-free networks
and for networks in which all $\idXi \leqslant b$, \mref{eq:decay} suggests that
a correspondingly large fraction of the local distributions we will estimate in
Algorithm \ref{algo:greedy} will have a small number $j$ of parents.
Furthermore, we find that in our experience BNs will typically have a weakly
connected DAG (that is, with no isolated nodes); and in this case local
distributions with $j = 0, 1$ will need to be learned for all nodes, and those
with $j = 2$ for all non-root nodes.

In the case of GBNs, local distributions for $j = 0, 1, 2$ parents can be
estimated in closed form using simple expressions as follows:
\begin{itemize}
  \item $j = 0$ corresponds to trivial linear regressions of the type
    \begin{equation*}
      X_i = \mu_{X_i} + \varepsilon_{X_i}.
    \end{equation*}
    in which the only parameters are the mean and the variance of $X_i$.
  \item $j = 1$ corresponds to simple linear regressions of the type
    \begin{equation*}
      X_i = \mu_{X_i} + X_j \beta_{X_j} + \varepsilon_{X_i},
    \end{equation*}
    for which there are the well-known \citep[\emph{e.g.}][]{draper}
    closed-form estimates
    \begin{align*}
      \hat{\mu}_{X_i} &= \bar{x}_i - \hat{\beta}_{X_j}\bar{x}_j, \\
      \hat{\beta}_{X_j} &= \frac{\COV(X_i, X_j)}{\VAR(X_i)}, \\
      \hat{\sigma}^2_{X_i} &= \frac{1}{n - 2}(X_i - \hat{x}_i)^T (X_i - \hat{x}_i);
    \end{align*}
    where $\VAR(\cdot)$ and $\COV(\cdot, \cdot)$ are empirical variances and
    covariances.
  \item for $j = 2$, we can estimate the parameters of
  \begin{equation*}
     X_i = \mu_{X_i} + X_j \beta_{X_j} + X_k \beta_{X_k} + \varepsilon_{X_i}
  \end{equation*}
  using their links to partial correlations:
  \begin{align*}
    \rho_{X_i X_j \given X_k}
      &= \frac{\rho_{X_i X_j} - \rho_{X_i X_k} \rho_{X_j X_k}}
             {\sqrt{1 - \rho_{X_i X_k}^2}\sqrt{1 - \rho_{X_j X_k}^2}} \\
      &= \beta_{j} \frac{\sqrt{1 - \rho_{X_j X_k}^2}}{\sqrt{1 - \rho_{X_i X_k}^2}}; \\
    \rho_{X_i X_k \given X_j}
      &= \beta_{k} \frac{\sqrt{1 - \rho_{X_j X_k}^2}}{\sqrt{1 - \rho_{X_i X_j}^2}};
  \end{align*}
  for further details we refer the reader to \citet{weatherburn}. Simplifying
  these expressions leads to
  \begin{align*}
    \hat{\beta}_{X_j} &= \frac{1}{d} \big[\VAR(X_k)\COV(X_i, X_j) - \\
      &\hspace{0.32\columnwidth} \COV(X_j, X_k)\COV(X_i, X_k)\big], \\
    \hat{\beta}_{X_k} &= \frac{1}{d} \big[\VAR(X_j)\COV(X_i, X_k) - \\
      &\hspace{0.32\columnwidth} \COV(X_j, X_k)\COV(X_i, X_j)\big];
  \end{align*}
  with denominator
  \begin{equation*}
    d = \VAR(X_j)\VAR(X_k) - \COV(X_j, X_k).
  \end{equation*}
  Then, the intercept and the standard error estimates can be computed as
  \begin{align*}
    \hat{\mu}_{X_i} &= \bar{x}_i - \hat{\beta}_{X_j}\bar{x}_j - \hat{\beta}_{X_k}\bar{x}_k, \\
    \hat{\sigma}^2_{X_i} &= \frac{1}{n - 3}(X_i - \hat{x}_i)^T (X_i - \hat{x}_i).
  \end{align*}
\end{itemize}

All these expressions are based on the variances and the covariances of
$(X_i, \PXi)$, and therefore can be computed in
\begin{multline}
  \underbrace{O\left(\frac{1}{2}n(1 + j)^2\right)}_{\text{covariance matrix of $(X_i, \PXi)$}} + \\
  \underbrace{O(n(1 + j))}_{\text{computing $\hat{x}_i$}} +
  \underbrace{O(3n)}_{\text{computing $\hat{\sigma}^2_{X_i}$}},
\label{eq:faster-gbn}
\end{multline}
This is lower than the computational complexity from \mref{eq:gbn-node} for the
same number of parents:
\begin{center}
\def\arraystretch{1.3}
\begin{tabular}{r|rr}
  j & from \mref{eq:gbn-node} & from \mref{eq:faster-gbn} \\
  \hline
  0 & $O(6n)$                 & $O(4.5n)$ \\
  1 & $O(9n)$                 & $O(7n)$\\
  2 & $O(16n)$                & $O(10.5n)$ \\
\end{tabular}
\end{center}
and it suggests that learning low-order local distributions in this way can be
markedly faster, thus driving down the overall computational complexity of
greedy search without any change in its behaviour. We also find that issues with
singularities and numeric stability, which are one of the reasons to use the QR
decomposition to estimate the regression coefficients, are easy to diagnose
using the variances and the covariances of $(X_i, \PXi)$; and they can be
resolved without increasing computational complexity again.

As for CLGBNs, similar reductions in complexity are possible for continuous
nodes. Firstly, if a continuous $X_i$ has no discrete parents
($\DXi = \varnothing$) then the computational complexity of learning its local
distribution using QR is again given by \mref{eq:gbn-node} as we noted in
Section \ref{sec:revisiting-clgbn-nodes}; and we are in the same setting we just
described for GBNs. Secondly, if $X_i$ has discrete parents ($D_{X_i} > 0$) and
$j$ continuous parents ($G_{X_i} = j$), the closed-form expressions above can
be computed for all the configurations of the discrete parents in
\begin{align}
  &\sum_{\dXi \in Val(D_{X_i})} O\left( \frac{1}{2}n_{\dXi}(1 +j)^2 \right) + \notag \\
  &\hspace{0.35\columnwidth} O(n_{\dXi}(1 + j)) + O(3n_{\dXi}) = \notag \\
  &\qquad = O\left( \frac{1}{2}n(1 + j)^2 \right) + O(n(1 + j)) + O(3n)
\label{eq:faster-clgbn}
\end{align}
time, which is lower than that required by the estimator from
\mref{eq:clgbn-node}:
\begin{center}
\def\arraystretch{1.3}
\begin{tabular}{r|rr}
  j & from \mref{eq:clgbn-node}         & from \mref{eq:faster-clgbn} \\
  \hline
  0 & $O\left(6n + l^{D_{X_i}}\right)$   & $O(4.5n)$ \\
  1 & $O\left(11n + 4l^{D_{X_i}}\right)$ & $O(7n)$\\
  2 & $O\left(18n + 9l^{D_{X_i}}\right)$ & $O(10.5n)$ \\
\end{tabular}
\end{center}
Interestingly we note that \mref{eq:faster-clgbn} does not depend on $D_{X_i}$,
unlike \mref{eq:clgbn-node}; the computational complexity of learning local
distributions with $G_{X_i} \leqslant 2$ does not become exponential even if
the number of discrete parents is not bounded.

\subsection{Predicting is Faster than Learning}
\label{sec:pred}

BNs are often implicitly formulated in a \emph{prequential setting} \citep{dawid},
in which a data set $\D$ is considered as a snapshot of a continuous stream of
observations and BNs are learned from that sample with a focus on predicting
future observations. \citet{engineering} called this the ``engineering
criterion'' and set
\begin{equation}
  \Score(\G, \D) = \log\Prob(\X^{(n+1)} \given \G, \Theta, \D)
\label{eq:eng}
\end{equation}
as the score function to select the optimal $\Gmax$, effectively maximising the
negative cross-entropy between the ``correct'' posterior distribution of
$\X^{(n + 1)}$ and that determined by the BN with DAG $\G$. They showed that
this score is consistent and that even for finite sample sizes it produces BNs
which are at least as good as the BNs learned using the scores in Section
\ref{sec:intro}, which focus on fitting $\D$ well. \citet{vanallen} and later
\citet{pena-cv} confirmed this fact by embedding $k$-fold cross-validation into
greedy search, and obtaining both better accuracy both in prediction and network
reconstruction. In both papers the use of cross-validation was motivated by the
need to make the best use of relatively small samples, for which the
computational complexity was not a crucial issue.

However, in a big data setting it is both faster and accurate to estimate
\mref{eq:eng} directly by splitting the data into a training and test set and
computing
\begin{equation}
  \Score(\G, \D) = \log\Prob(\D^{test} \given \G, \Theta, \D^{train});
\label{eq:eng2}
\end{equation}
that is, we learn the local distributions on $\D^{train}$ and we estimate the
probability of $\D^{test}$. As is the case for many other models
\citep[\emph{e.g.}, deep neural networks;][]{benjo}, we note that prediction is
computationally much cheaper than learning because it does not involve solving
an optimisation problem. In the case of BNs, computing
\mref{eq:eng2} is:
\begin{itemize}
  \item $O(N|\D^{test}|)$ for discrete BNs, because we just have to perform an
    $O(1)$ look-up to collect the relevant conditional probability for each node
    and observation;
  \item $O(cN|\D^{test}|)$ for GBNs and CLGBNs, because for each node and
    observation we need to compute \newline $\PXi^{(n+1)}\hat{\beta}_{X_i}$ and
    $\hat{\beta}_{X_i}$ is a vector of length $\idXi$.
\end{itemize}
In contrast, using the same number of observations for learning in GBNs and
CLGBNs involves a QR decomposition to estimate the regression coefficients of
each node in both \mref{eq:gbn-node} and \mref{eq:clgbn-node}; and that takes
longer than linear time in $N$.

\begin{figure*}[t]
  \begin{center}
  \includegraphics[width=0.8\linewidth]{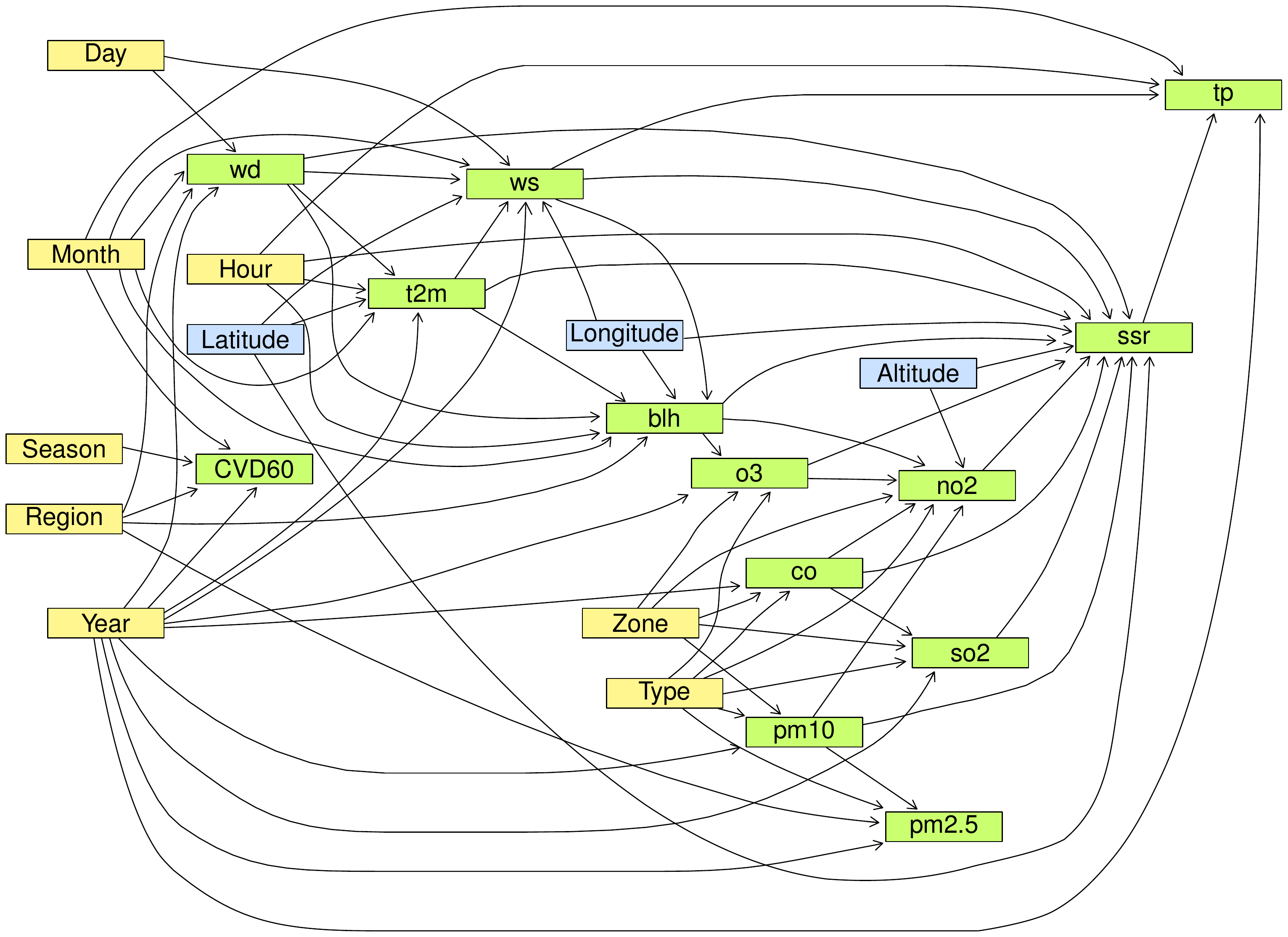}
  \caption{Conditional Linear Gaussian BN from \citet{vitolo17}. Yellow nodes are
    multinomial, blue nodes are Gaussian, and green nodes are conditional linear
    Gaussian.}
  \label{fig:bn}
  \end{center}
\end{figure*}

\noindent
Hence by learning local distributions only on $\D^{train}$ we lower the overall
computational complexity of structure learning because the per-observation cost
of prediction is lower than that of learning; and $\D^{train}$ will still be
large enough to obtain good estimates of their parameters $\T$. Clearly, the
reduction in complexity will be determined by the proportion of $\D$ used as
$\D^{test}$. Further speed-ups are possible by using the closed-form results
from Section \ref{sec:closed} to reduce the complexity of learning local
distributions on $\D^{train}$, combining the effect of all the optimisations
proposed in this section.

\section{Benchmarking and Simulations}
\label{sec:sims}

We demonstrate the reductions in computational complexity we discussed in
Sections \ref{sec:closed} and \ref{sec:pred} using the MEHRA data set from
\citet{vitolo17}, which studied 50 million observations to explore the
interplay between environmental factors, exposure levels to outdoor air
pollutants, and health outcomes in the English regions of the United Kingdom
between 1981 and 2014. The CLGBN learned in that paper is shown in Figure
\ref{fig:bn}: it comprises $24$ variables describing the concentrations of
various air pollutants (O3, PM\textsubscript{2.5}, PM\textsubscript{10},
SO\textsubscript{2}, NO\textsubscript{2}, CO) measured in $162$ monitoring
stations, their geographical characteristics (latitude, longitude, latitude,
region and zone type), weather (wind speed and direction, temperature, rainfall,
solar radiation, boundary layer height), demography and mortality rates.

The original analysis was performed with the \emph{bnlearn} R package
\citep{jss09}, and it was complicated by the fact that many of the variables
describing the pollutants had significant amounts of missing data due to the
lack of coverage in particular regions and years. Therefore, \citet{vitolo17}
learned the BN using the Structural EM algorithm \citep{sem}, which is an
application of the Expectation-Maximisation algorithm \citep[EM;][]{em} to BN
structure learning that uses hill-climbing to implement the M step.

For the purpose of this paper, and to better illustrate the performance
improvements arising from the optimisations from Section \ref{sec:bigdata},
we will generate large samples from the CLGBN learned by \citet{vitolo17} to be
able to control sample size and to work with plain hill-climbing on complete
data. In particular:
\begin{enumerate}
  \item we consider sample sizes of $1$, $2$, $5$, $10$, $20$ and $50$ millions;
  \item for each sample size, we generate $5$ data sets from the CLGBN;
  \item for each sample, we learn back the structure of the BN using
    hill-climbing using various optimisations:
    \begin{itemize}
      \item QR: estimating all Gaussian and conditional linear Gaussian local
        distributions using the QR decomposition, and BIC as the score function;
      \item 1P: using the closed form estimates for the local distributions that
        involve $0$ or $1$ parents, and BIC as the score function;
      \item 2P: using the closed form estimates for the local distributions that
        involve $0$, $1$ or $2$ parents, and BIC as the score functions;
      \item PRED: using the closed form estimates for the local distributions
        that involve $0$, $1$ or $2$ parents for learning the local
        distributions on $75\%$ of the data and estimating \mref{eq:eng2} on
        the remaining $25\%$.
    \end{itemize}
\end{enumerate}
For each sample and optimisation, we run hill-climbing $5$ times and we average
the resulting running times to reduce the variability of each estimate.
Furthermore, we measure the accuracy of network reconstruction using the
Structural Hamming Distance \citet[SHD;][]{mmhc}, which measures the number of
arcs that differ between the CPDAG representations of the equivalence classes
of two network structures. In our case, those we learn from the simulated data
and the original network structure from \citet{vitolo17}. All computations are
performed with the \emph{bnlearn} package in R 3.3.3 on a machine with two Intel
Xeon CPU E5-2690 processors (16 cores) and 384GB of RAM.

\begin{figure}[t]
  \begin{center}
  \includegraphics[width=\linewidth]{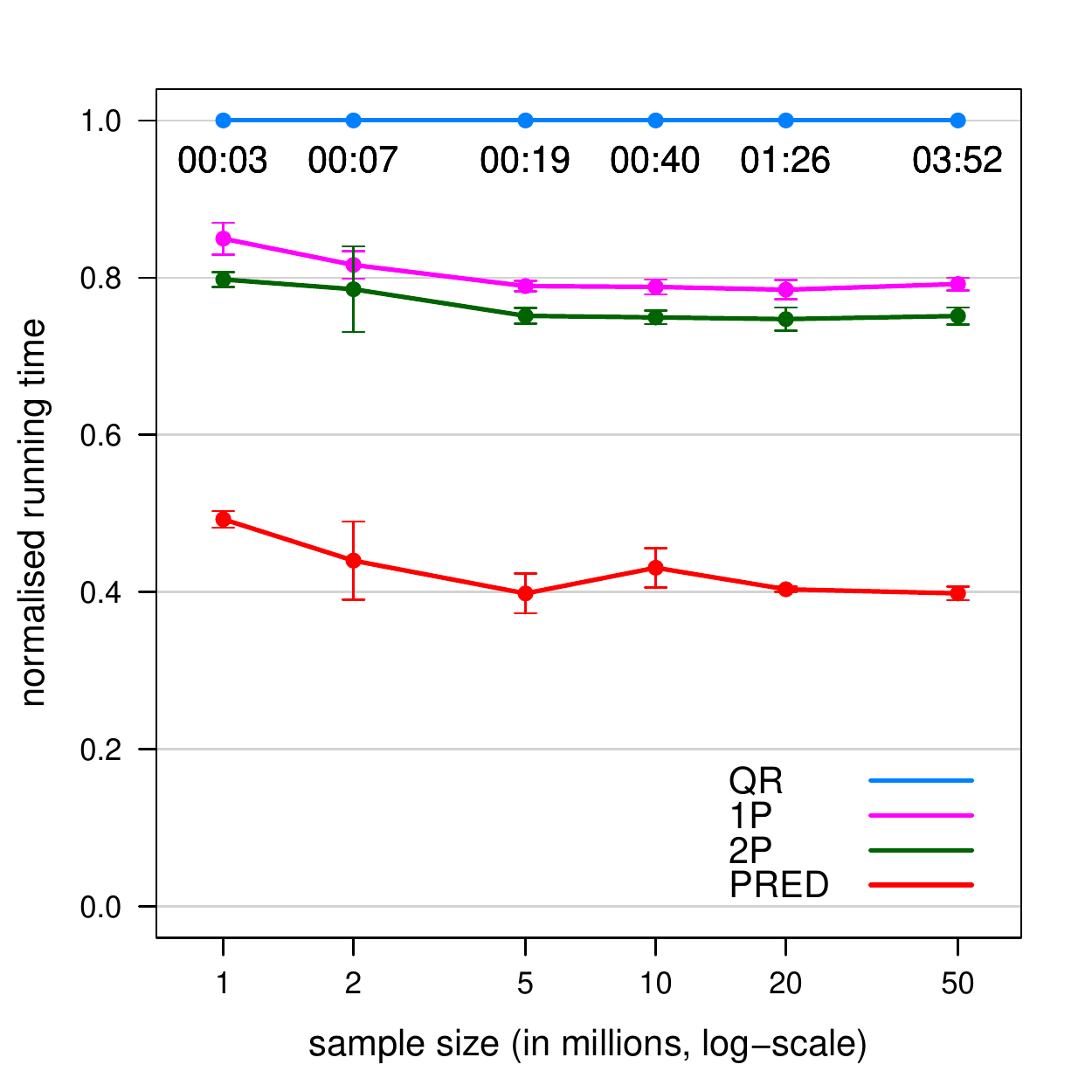}
  \caption{Running times for the MEHRA data set, normalised using the baseline
    implementation based on the QR decomposition (blue), for 1P (pink), 2P
    (green) and PRED (red). Bars  represent 95\% confidence intervals. Average
    running times are reported for QR.}
  \label{fig:bench-claudia}
  \end{center}
\end{figure}

The running times for 1P, 2P and PRED, normalised using those for QR as a
baseline, are shown in Figure \ref{fig:bench-claudia}. As expected,
computational complexity gradually decreases with the level of optimisation: 1P
(pink) is $\approx 20\%$ faster than QR, 2P (green) is $\approx 25\%$ faster and
PRED (red) is $\approx 60\%$ faster, with minor variations at different sample
sizes. PRED exhibits a larger variability because of the randomness introduced
by the subsampling of $\D^{test}$, and provides smaller speed-ups for the
smallest considered sample size ($1$ million). Furthermore, we confirm the
results from \citet{engineering} on network reconstruction accuracy. In Table
\ref{tab:shd-mehra} we report the sums of the SHDs between the network
structures learned by BIC and that from \citet{vitolo17}, and the corresponding
sum for the networks learned using PRED, for the considered sample sizes.
Overall, we find that BIC results in $13$ errors over the $30$ learned DAGs,
compared to $4$ for \mref{eq:eng2}. The difference is quite marked for samples
of size 1 million (11 errors versus 2 errors). On the other hand, neither score
results in any error for samples with more than 10 million observations, thus
confirming the consistency of PRED. Finally we confirm that the observed running
times increase linearly in the sample size as we showed in Section
\ref{sec:revisiting}.

\begin{table}[t]
  \begin{center}
  \def\arraystretch{1.3}
  \begin{tabular}{r|rr}
    n  & BIC & PRED \\
    \hline
    1  & 11  & 2    \\
    2  & 2   & 1    \\
    5  & 0   & 1    \\
    10 & 0   & 0    \\
    20 & 0   & 0    \\
    50 & 0   & 0
  \end{tabular}
  \caption{Sums of the SHDs between the network structures learned by BIC, PRED
    and that from \citet{vitolo17} for different sample sizes $n$.}
  \label{tab:shd-mehra}
  \end{center}
\end{table}

\begin{table}[b]
  \begin{center}
  \def\arraystretch{1.3}
  \begin{tabular}{l|rrr|p{3cm}}
    Data    & $n$          & $d$ & $M$ & reference \\
    \hline
    AIRLINE & $53.6 \times 10^6$ & $9$ & $19$ & \citet{airline} \\
    GAS     & $ 4.2 \times 10^6$ & $0$ & $37$ & UCI ML Repository, \citet{gas} \\
    HEPMASS & $10.5 \times 10^6$ & $1$ & $28$ & UCI ML Repository, \citet{hepmass} \\
    HIGGS   & $11.0 \times 10^6$ & $1$ & $28$ & UCI ML Repository, \citet{higgs} \\
    SUSY    & $ 5.0 \times 10^6$ & $1$ & $18$ & UCI ML Repository, \citet{higgs} \\
  \end{tabular}
  \caption{Data sets from the UCI Machine Learning Repository and the JSM Data
    Exposition session, with their sample size ($n$), multinomial nodes ($N - M$)
    and Gaussian/conditional Gaussian nodes ($M$).}
  \label{tab:uci}
  \end{center}
\end{table}

\begin{figure}[t]
  \begin{center}
  \includegraphics[width=\linewidth]{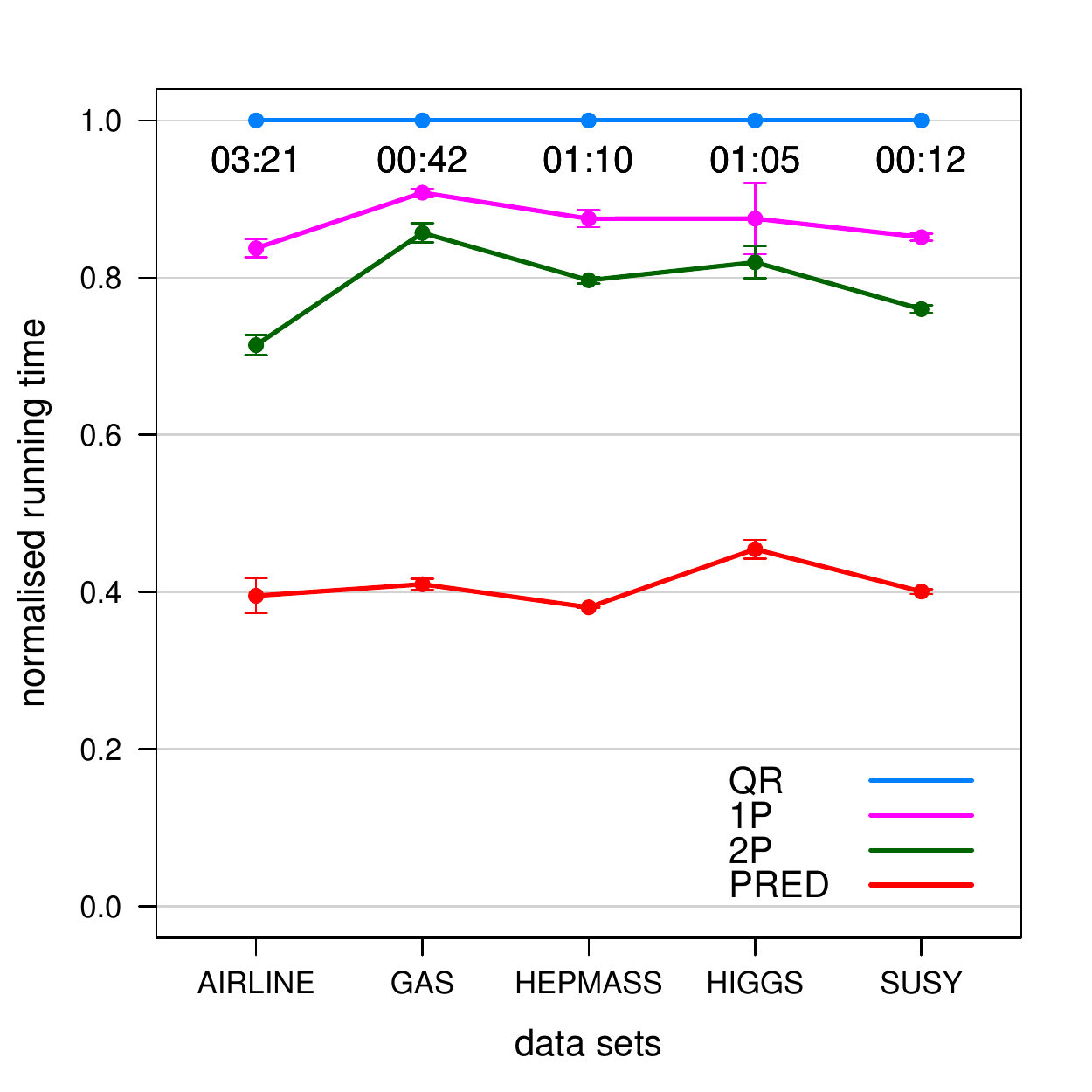}
  \caption{Running times for the data sets in Table \ref{tab:uci}, normalised
    using the baseline implementation based on the QR decomposition (blue), for
    1P (pink), 2P (green) and PRED (red). Bars represent 95\% confidence
    intervals. Average running times are reported for QR.}
  \label{fig:bench-uci}
  \end{center}
\end{figure}

In order to verify that these speed increases extend beyond the MEHRA data set,
we considered five other data sets from the UCI Machine Learning Repository
\citep{uci} and from the repository of the Data Exposition Session of the
Joint Statistical Meetings (JSM). These particular data sets have been chosen
because of their large sample sizes and because they have similar
characteristics to MEHRA (continuous variables, a few discrete variables,
20-40 nodes overall; see Table \ref{tab:uci} for details). However, since their
underlying ``true DAGs'' are unknown, we cannot comment on the accuracy of the
DAGs we learn from them. For the same reason, we limit the density of the
learned DAGs by restricting each node to have at most 5 parents; this produces
DAGs with $2.5N$ to $3.5N$ arcs depending on the data set. The times for 1P, 2P
and PRED, again normalised by those for QR, are shown in Figure
\ref{fig:bench-uci}. Overall, we confirm that PRED is $\approx 60\%$ faster on
average than QR. Compared to MEHRA, 1P and 2P are to some extend slower with
average speed-ups of only $\approx 15\%$ and $\approx 22\%$ respectively.
However, it is apparent by comparing Figures \ref{fig:bench-claudia} and
\ref{fig:bench-uci} that the reductions in computational complexity are
consistent over all the data sets considered in this paper, and hold for a
wide range of sample sizes and combinations of discrete and continuous
variables.

\section{Conclusions}
\label{sec:conclusions}

Learning the structure of BNs from large data sets is a computationally
challenging problem. After deriving the computational complexity of the greedy
search algorithm in closed form for discrete, Gaussian and conditional linear
Gaussian BNs, we studied the implications of the resulting expressions in a
``big data'' setting where the sample size is very large, and much larger than
the number of nodes in the BN. We found that, contrary to classic
characterisations, computational complexity strongly depends on the class of BN
being learned in addition to the sparsity of the underlying DAG. Starting from
this result, we suggested two possible optimisations to lower the computational
complexity of greedy search and thus speed up the most common algorithm used for
BN structure learning. Using a large environmental data set and five data sets
from the UCI Machine Learning Repository and the JSM Data Exposition, we show
that it is possible to reduce the running time greedy search by
$\approx 60\%$.


\end{document}